\documentstyle [12pt] {article}
\begin{document}

\def\ri{\rightarrow}
\def\L{{\cal L}}
\def\D{{\buildrel\leftarrow\over D\!\!\!\!\!/}}
\def\lem{{\cal L}_{\rm em}}
\def\half{{1\over 2}}
\def\shalf{{1\over \sqrt{2}}}
\def\xip{{\Xi'}}
\def\es{\varepsilon}
\def\pc{{\rm pc}}
\def\up{\uparrow}
\def\dw{\downarrow}
\def\Q{{\cal Q}}
\def\ra{\rangle}
\def\la{\langle}
\def\m{{m_{_{\Xi_c}}\over m_{_{\Xi_b}}}}
\def\mm{{m_{_{\Xi_b}}\over m_{_{\Xi_c}}}}
\def\pr{{\sl Phys. Rev.}~}
\def\prl{{\sl Phys. Rev. Lett.}~}
\def\pl{{\sl Phys. Lett.}~}
\def\np{{\sl Nucl. Phys.}~}
\def\zp{{\sl Z. Phys.}~}

\font\el=cmbx10 scaled \magstep2
{\obeylines
\hfill CLNS 94/1278
\hfill IP-ASTP-04-94}

\vskip 0.7 cm

\centerline {{\el Effective Lagrangian Approach to}}
\centerline{{\el Weak Radiative Decays of Heavy Hadrons}}

\medskip
\bigskip

\centerline{\bf Hai-Yang Cheng$^{a}$, Chi-Yee Cheung$^a$, Guey-Lin Lin$
^{a,\dagger}$,}
\centerline{\bf Y.C. Lin$^b$, Tung-Mow Yan$^{c}$, and Hoi-Lai Yu$^a$}

\medskip
\centerline{$^a$ Institute of Physics, Academia Sinica, Taipei,}
\centerline{Taiwan 11529, Republic of China}

\medskip
\centerline{$^b$ Physics Department, National Central University,
Chung-li,}
\centerline{Taiwan 32054, Republic of China}

\medskip
\centerline{$^c$ Floyd R. Newman Laboratory of Nuclear Studies, Cornell
University}
\centerline{Ithaca, New York 14853, USA}
\bigskip
\bigskip

\centerline{\bf Abstract}

Motivated by the observation of the decay $\bar{B}\to \bar{K}^*\gamma$ by
CLEO, we have systematically analyzed the two-body weak radiative decays of
bottom and charmed hadrons. There exist
two types of weak radiative decays: One proceeds through the short-distance
$b\to s\gamma$ transition  and the other
occurs through $W$-exchange accompanied by a photon emission. Effective
Lagrangians are derived for the $W$-exchange bremsstrahlung processes
at the quark level and then applied to various weak electromagnetic decays
of heavy hadrons. Predictions for the branching ratios of $\bar{B}^0\to D^{*0}
\gamma,~\Lambda_b^0\to\Sigma_c^0\gamma,~\Xi_b^0\to
\Xi_c^0\gamma$ and $\Xi_b^0\to\xip_c^0\gamma$
are given. In particular, we found ${\cal B}(\bar{B}^0
\to D^{*0}\gamma)\approx 0.9\times 10^{-6}$. Order of magnitude estimates
for the weak radiative decays of charmed hadrons:
$~D^0\to \bar{K}^{*0}\gamma,~\Lambda_c^+\to\Sigma^+\gamma$ and
$\Xi_c^0\to\Xi^0\gamma$ are also presented. Within this
approach, the decay asymmetry for antitriplet to antitriplet heavy baryon
weak radiative transitions is uniquely predicted by heavy quark symmetry.
The electromagnetic penguin contribution to $\Lambda_b^0\to\Lambda\gamma$
is estimated by two different methods and its branching ratio is found to
be of order $1\times 10^{-5}$. We conclude that weak
radiative decays of bottom hadrons are dominated by the short-distance
$b\to s\gamma$ mechanism.

\pagebreak

\noindent {\bf I.~~Introduction}
\vskip 0.4cm
 The recent observation of the weak radiative decay ${\bar B}\ri \bar{K}^*
\gamma$ by CLEO [1] with the branching ratio $(4.5\pm 1.5\pm 0.9)\times 10^
{-5}$ confirms the standard-model expectation that this decay mode is
dominated by the short-distance electromagnetic penguin transition $b\ri s
\gamma$. Naively, it is tempting to think that $\bar{B}\ri D^*\gamma$ will be
the dominant weak radiative
decay of the $\bar{B}$ meson as it is not suppressed by quark mixing angles.
However, owing to the large top quark mass, the amplitude of
$b\ri s\gamma$ is neither quark mixing nor loop suppressed. Moreover, it is
largely enhanced by QCD corrections. As a consequence, the short-distance
contribution due to the electromagnetic penguin diagram
dominates over the $W$-exchange bremsstrahlung.
This phenomenon is quite unique to the bottom hadrons which contain a heavy
$b$ quark; such a magic short-distance enhancement does not occur in the
systems of charmed and strange hadrons. For example, it is known that the
mechanism $s\ri d\gamma$ plays only a minor role in the radiative decays of
kaons and hyperons.

   In Ref.[2] we have systematically studied the flavor-conserving
electromagnetic decays of heavy mesons and heavy baryons. Various photon
coupling constants are related through the usage of heavy quark symmetry. For
example, the $\bar{B}^*\bar{B}^*\gamma$ coupling, which is very difficult to
measure in practice, is related to the $\bar{B}^*\bar{B}\gamma$ coupling via
heavy-quark spin
symmetry. The coupling constants appearing in the Lagrangians depend only on
the light quarks and can be calculated in the nonrelativistic quark model.
Consequently, the dynamics of the electromagnetic transitions for emissions
of soft photons and pions is completely determined by heavy quark and chiral
symmetry, supplemented by the quark model. The purpose of the present paper
is to extend our previous work to the weak radiative decays of heavy hadrons.

   At the quark level, there are three different types of processes which can
contribute to the weak radiative decays of heavy hadrons, namely, single-,
two- and three-quark transitions [3]. The single-quark transition
mechanism comes from the so-called electromagnetic penguin diagram. Since
the penguin process $c\to u\gamma$ is very suppressed, it plays no role in
charmed hadron radiative decays. We will thus focus on the
two-body radiative decays of bottom hadrons proceeding through
the electromagnetic penguin mechanism $b\ri s\gamma$:
$$
\begin{array}{ccl}
 && \bar{B} \ri  \bar{K}^*\gamma,~~~\bar{B}_s\ri\phi\gamma,  \\
 &&\Lambda_b^0 \ri \Sigma^0\gamma,~\Lambda^0\gamma,~~~\Xi^0_b\ri\Xi^0\gamma,
{}~~~\Xi^-_b\ri\Xi^-\gamma,~~~\Omega_b^-\to\Omega^-\gamma.
\end{array}\eqno(1.1)
$$
There are two contributions from the two-quark transitions: one from the
$W$-exchange diagram accompanied by a photon emission from the external
quark (see, for example, Fig. 1), and the other from the same $W$-exchange
diagram but with
 a photon radiated from the $W$ boson. The latter is typically suppressed by
a factor of $m_qk/M_W^2$ ($k$ being the photon energy) as compared to the
former bremsstrahlung process [4]. For bottom hadrons, the dominant decays
which occur through the quark-quark bremsstrahlung $b\bar{d}\to c\bar{u}
\gamma$ or $bu\to cd\gamma$ are:
$$
\begin{array}{ccl}
 && \bar{B}^0 \ri D^{*0}\gamma, \\
 && \Lambda_b^0  \ri \Sigma^0_c\gamma,~~~\Xi^0_b\ri\Xi^0_c\gamma,~{\Xi'}^0_c
\gamma,
\end{array}\eqno(1.2)
$$
where we have followed the convention that a $\bar{B}$ meson contains a $b$
quark and that $\Xi_Q~(\Xi'_Q)$ denote antitriplet (sextet) heavy baryons.
For charmed hadrons, the Cabibbo-allowed decay modes via $c\bar{u}\to
s\bar{d}\gamma$ or $cd\to us\gamma$ are
$$\begin{array}{ccl}
  && D^0\to \bar{K}^{*0}\gamma,  \\  &&\Lambda_c^+\to\Sigma^+\gamma,~~~~
\Xi_c^0\to\Xi^0\gamma.
\end{array}\eqno(1.3)$$
Note that some decay modes in (1.1) also receive contributions from
$W$-exchange bremsstrahlung, but they are suppressed by quark mixing angles.
Finally, the three-quark transition involving $W$-exchange between two quarks
and a photon emission by the third quark is quite suppressed because of very
small probability of finding three quarks in adequate kinematic matching
with the baryons [3,5].

   To summarize, the two important mechanisms for weak radiative decays of
heavy hadrons are $W$-exchange bremsstrahlung and the electromagnectic
penguin transition $b\to s\gamma$. Since the effective Lagrangian for the
latter is known, the calculation for the radiative amplitude induced by
the penguin diagram appears easier at first sight.

  The $W$-exchange bremsstrahlung effect is usually evaluated under the pole
assumption; that is, its amplitude is saturated by one-particle intermediate
states. When dealing with weak radiative decays of heavy hadrons, one
encounters two predicaments. First,
the hadronic matrix elements for the processes (1.1) are evaluated at $q^2=0$
for a real photon emission, whereas heavy quark symmetry and the quark model
are known to be more reliable at zero recoil kinematic point where $q^2$ is
maximum. (The quark-model wave functions best resemble the hadron states in
the frame where both hadrons are {\it static}.) Second, the intermediate states
appearing in the pole diagrams for the processes (1.2) or (1.3) are very far
from their mass
shell. For example, the four-momentum squared of the $D$ pole in the decay
$\bar{B}\ri D^*\gamma$ is $m^2_B$ (see Fig. 3). This means that the residual
momentum of the $D$ meson defined by $P_\mu=m_{_D}v_\mu+k_\mu$ must be of
order $m_B$, so the approximation $k/m_{_D}<<1$ required by the heavy quark
effective theory is no longer valid. Also, the quark-model prediction for the
photon coupling constants is presumably reliable only when both hadrons are
nearly on their mass shell. The
question is then how to extrapolate the hadronic matrix elements from zero
recoil to maximal recoil, and the photon couplings from the on-shell point to
off-shell? In principle, one can treat the intermediate state as an on-shell
particle and then
assume that off-shell effects of the pole can be parametrized in terms of
form factors. Such form factors are basically unknown, though they are
expected to become smaller as the
intermediate state is more away from its mass shell due to less overlap
of initial and final hadron wave functions. Consequently, based on heavy quark
symmetry and the nonrelativistic quark model, at best we can only
predict the upper bound of the decay rates for the radiative decays in the
category of (1.2).

 We will present in this  paper a different but more powerful approach to the
$W$-exchange bremsstrahlung processes. The fact that the intermediate quark
state in these processes is sufficiently off-shell (see, e.g. Fig. 1) and
the emitted photon is hard suggests the possibility of analyzing
these processes by perturbative QCD. As a first step in this direction, we
study the tree amplitudes responsible for these processes and derive a gauge
invariant effective five-point interaction for the quark-quark bremsstrahlung
$bu\to cd\gamma$ or $b\bar{d}\to c\bar{u}\gamma$.

    The physical mass of a heavy hadron differs from the heavy quark mass by
an amount of order $\Lambda_{\rm QCD}$. This difference is due to the presence
of the light quark(s). It is therefore reasonable to assign a constitutent
mass of order $\Lambda_{\rm QCD}$ to the light quark(s) inside a heavy hadron.
In addition, the light quarks move, on average, with the same velocity as
the heavy quark. We will make the simplifying assumption of neglecting the
relative Fermi motion. Thus, the heavy quark and the light quark(s) in a heavy
hadron move with equal four velocity. This momentum parametrization has the
advantage that the resulting effective interaction is local and manifestly
gauge invariant.
In Sec. III we will show explicitly for the
meson case that the effective Lagrangian and the pole model
approaches are indeed equivalent, but the former is much simpler and provides
information on the form factors.

   Armed with the effective Lagrangian for the $W$-exchange bremsstrahlung,
we are able to study various radiative decay modes of bottom and charmed
hadrons listed in (1.2) and (1.3), bearing in mind that this approach
presumably works better when both the initial and final hadrons contain a
heavy quark. We will use the factorization method, which is known to
work well for nonleptonic weak decays of heavy mesons, to evaluate the
mesonic matrix elements. As for the baryon radiative decays, we will
demonstrate that heavy quark symmetry leads to a nontrivial prediction for
 antitriplet to antitriplet heavy baryon transitions: The ratio of
parity-conserving and parity-violating amplitudes is uniquely predicted.
Baryonic matrix elements will be calculated using the MIT bag model.

   This paper is organized as follows. Local effective Lagrangians for the
quark-quark bremsstrahlung processes are derived in Sec. II. We then apply
this approach to weak radiative decays of heavy mesons in Sec. III and to
various bottom decays in Sec. IV.
Discussion and conclusion are presented in Section V.
Some preliminary results have been reported earlier [6].

\pagebreak
\noindent {\bf II.~~Effective Lagrangians for Weak Radiative Decays}
\vskip 0.3 cm
   In this Section we will present the effective Lagrangians for the penguin
transition $b\to s+\gamma$ and for the $W$-exchange bremsstrahlung processes
$b+u\to c+d+\gamma$ and $b+\bar{d}\to c+\bar{u}+\gamma$.

 The effective Lagrangian for the short-distance $b\ri s\gamma$ transition
including QCD corrections reads [7]
$${\cal L}_{\rm eff}(b\ri s\gamma)=\,{G_F\over 2\sqrt{2}}\,{e\over 8\pi^2}F_2
V_{tb}V_{ts}^*\,\bar{s}\sigma\cdot F[m_b(1+\gamma_5)+m_s(1-
\gamma_5)]b,\eqno(2.1)$$
where $\sigma\cdot F\equiv\sigma_{\mu\nu}F^{\mu\nu}$, and
we have neglected contributions which vanish for a real photon emission,
$V_{ij}$ is the quark mixing matrix element,
$F_2\cong F_2(x_t)-F_2(x_c)\cong F_2(x_t)$ with $x_i=m^2_i/M_W^2$, and
$$
F_2(x)=\,\rho^{-{16\over 23}}\left\{ \bar{F}_2(x)+{116\over 27}\left[{1\over 5}
\left(\rho^{10\over 23}-1\right)+{1\over 14}\left(\rho^{28\over 23}-1\right)
\right]\right\},\eqno(2.2)$$
with
$$
\bar{F}_2(x)  =  {(8x^2+5x-7)x\over 12(x-1)^3}-{(3x-2)x^2\over 2(x-1)
^4}\ln x,\eqno(2.3a)$$
$$\rho =  {\alpha_s(m^2_b)\over \alpha_s(M^2_W)}=1+{23\over 12
\pi}\alpha_s(m^2_b)\ln\left({M^2_W\over m_b^2}\right).\eqno(2.3b)
$$
It is easily seen that $F_2$ is a smooth function of the top quark mass. For
$m_t=150$ GeV and $\Lambda_{\rm QCD}=200$ MeV, we find
$\bar{F}_2(x_t)=0.34$ and $F_2(x_t)=0.73$, so that the radiative
decay $b\ri s\gamma$ is enhanced by QCD corrections by a factor of 2.
The radiative decays $\bar{B}\to\bar{K}^*\gamma$ and $\bar{B}\to \phi\gamma$,
mediated by this penguin mechanism, have been studied extensively in the
literature. In Section IV we will apply the effective Lagrangian (2.1) to the
decay $\Lambda_b\to \Lambda\gamma$.

    We next turn to the $W$-exchange bremsstrahlung processes $b\bar{d}\to c
\bar{u}\gamma$ and $bu\to cd\gamma$. The difficulty associated with highly
off-shell intermediate states mentioned in the Introduction is easily overcome
at the quark level. The propagator of the highly virtual quark can be reduced
to a constant by energy-momentum conservation.
The above photon emission reactions are then described by an effective
five-point local interaction which
is also gauge invariant. To begin with, we note that the relevant QCD-corrected
effective weak Hamiltonian is given by
$${\cal H}_{\rm eff}=\,{G_F\over 2\sqrt{2}}V_{cb}V^*_{ud}(c_+O_++c_-O_-)
+h.c.~,\eqno(2.4)$$
with
$$O_\pm=\,O_A\pm O_B\eqno(2.5)$$
and
$$O_A=\,(\bar{c}b)(\bar{d}u),~~~~O_B=\,(\bar{c}u)(\bar{d}b),\eqno(2.6)$$
where $(\bar{q}_1q_2)\equiv\bar{q}_1\gamma_\mu(1-\gamma_5)q_2$. The Wilson
coefficient functions $c_\pm$, evaluated at the scale $\mu= m_b$, have the
values
$$c_+(m_b)\simeq 0.85\,,~~~~c_-(m_b)\simeq 1.38~.\eqno(2.7)$$

We first consider the photon emission process $b\bar{d}\to c\bar{u}
\gamma$. The amplitudes mediated by the operator $O_A$ (see Fig. 1) are
$$A_1=\,ee_c\bar{u}_c\gamma^\mu{1\over p\!\!\!/ _c+k\!\!\!/ -m_c}\gamma^\nu
(1-\gamma_5)u_b\bar{v}_d\gamma_\nu(1-\gamma_5)v_u,\eqno(2.8a)$$
$$A_2=\,ee_b\bar{u}_c\gamma^\nu(1-\gamma_5){1\over p\!\!\!/ _b-k\!\!\!/ -m_b}
\gamma^\mu u_b\bar{v}_d\gamma_\nu(1-\gamma_5)v_u,\eqno(2.8b)$$
$$A_3=\,ee_d\bar{u}_c\gamma^\nu(1-\gamma_5)u_b\bar{v}_d\gamma^\mu{1\over
-p\!\!\!/ _{\bar{d}}+k\!\!\!/ -m_d}\gamma_\nu(1-\gamma_5) v_u,\eqno(2.8c)$$
$$A_4=\,ee_u\bar{u}_c\gamma^\nu(1-\gamma_5)u_b\bar{v}_d\gamma_\nu(1-\gamma_5)
{1\over -p\!\!\!/ _{\bar{u}}-k\!\!\!/ -m_u}\gamma^\mu v_u,\eqno(2.8d)$$
where $k$ is the photon momentum.

  We will parametrize the quark momenta in terms of velocities; this is more
suitable when dealing with heavy quark symmetry:
$$p_b=m_bv,~~~p_{\bar d}=m_dv,~~~p_c=m_cv',~~~p_{\bar u}=m_uv'.\eqno(2.9)$$
Since the $b$ quark is heavy, we have set the velocity of the $\bar{d}$
quark to be the same as the $b$ quark so that they will move together to form
a bound meson state. Likewise, $v_{\bar u}=v_c=v'$. At this point we
wish to emphasize that the light quark masses appearing in (2.8)-(2.9) are
of the
constituent type. This is attributed to the fact the typical Fermi momentum of
the quarks in a hadron is of order $\Lambda_{\rm QCD}$. Consequently, although
the current quark masses of the light $u$ and $d$ quarks are only of order 10
MeV, their off-shellness is of order $\Lambda_{\rm QCD}$. We thus choose to
have the light quarks close to their mass shell, so that $p_q\approx m_qv$ with
$v^2=1$ and $m_q$ being the constituent quark mass. Obviously, this
parametrization (2.9) does not provide a complete description of the Fermi
motion inside the bound state. Nevertheless, it does take into account its
average effect by giving a constitutent mass of order $\Lambda_{\rm QCD}$
to the light quarks. This parametrization greatly simplifies the
calculation by eliminating the photon's coupling to the convection currents
and making the effective interaction local and manifestly gauge invariant.

   With the momentum parametrization given by (2.9), it is easily seen that
the contributions from the convection current add up to zero
and the amplitude arises entirely from the magnetic moments of the quarks:
$$
\begin{array}{ccl}
A &=& A_1+A_2+A_3+A_4 \\
  &=& -{ie\over m_i^2-m_f^2}\Big[\,e_c
{m_f\over m_c}\,\bar{u}_c\sigma^{\mu\lambda}k_\lambda \gamma^\nu(1-\gamma_5)
u_b\,\bar{v}_d\gamma_\nu(1-\gamma_5)v_u  \\
&-& e_b{m_i\over m_b}\,\bar{u}_c\gamma^\nu(1-\gamma_5)\sigma^{\mu\lambda}k_
\lambda u_b\,\bar{v}_d\gamma_\nu(1-\gamma_5)v_u  \\
&-& e_d{m_i\over m_d}\,\bar{u}_c\gamma^\nu(1-\gamma_5)u_b\,\bar{v}_d\sigma^{\mu
\lambda}k_\lambda\gamma_\nu(1-\gamma_5)v_u  \\
&+& e_u{m_f\over m_u}\,\bar{u}_c\gamma^\nu(1-\gamma_5)u_b\,\bar{v}_d\gamma_\nu
(1-\gamma_5)\sigma^{\mu\lambda}k_\lambda v_u~\Big],
\end{array}\eqno(2.10)$$
where $m_i=m_b+m_d$ and $m_f=m_c+m_u$. The amplitude corresponding to Fig.2
induced by the operator $O_B$ can be obtained from Eq.(2.10) by the
substitution $u_b\leftrightarrow v_u$.
The above amplitudes can be further
simplified by considering the commutator and anticommutator relations
$${1\over 2}\left\{ \sigma_{\mu\nu},~\gamma_\lambda\right\}=\,\epsilon_{\mu\nu
\lambda\alpha}\gamma^\alpha\gamma_5,\eqno(2.11a)$$
$${1\over 2}[\,\sigma_{\mu\nu},~\gamma_\lambda\,]=\,-i(g_{\mu\lambda}\gamma
_\nu-g_{\nu\lambda}\gamma_\mu).\eqno(2.11b)$$
We find that for the
photon emission process $b\bar{d}\to c\bar{u}\gamma$, we can simply replace
the operator $O_\pm$ in (2.4) by ${O}^F_\pm$
so that the effective Hamiltonian is given by
$${\cal H}_{\rm eff}(b\bar{d}\to c\bar{u}\gamma)=\,{G_F\over 2\sqrt{2}}V_{cd}
V_{ud}^*(c_+{O}^F_++c_-{O}^F_-),\eqno(2.12)$$
with
$$\begin{array}{ccl}
{O}^F_\pm(b\bar{d}\to c\bar{u}\gamma)=\,{e\over m_i^2-m_f^2} &\Big\{& \left(
e_c{m_f\over m_c}+e_d{m_i\over m_d}\right)\left(\tilde{F}_{\mu\nu}+iF_{\mu\nu}
\right)O_\pm^{\mu\nu}   \\
&-&\left(e_u{m_f\over m_u}+e_b{m_i\over m_b}\right)\left(\tilde{F}_{\mu\nu}
-iF_{\mu\nu}\right)O_\mp^{\mu\nu}\Big\},
\end{array}\eqno(2.13)$$
where
$$O_\pm^{\mu\nu}=\,\bar{c}\gamma^\mu(1-\gamma_5)b\bar{d}\gamma^\nu(1-\gamma_5)u
\pm \bar{c}\gamma^\mu(1-\gamma_5)u\bar{d}\gamma^\nu(1-\gamma_5)b,\eqno(2.14a)$$
$$\tilde{F}_{\mu\nu}\equiv {1\over 2}\epsilon_{\mu\nu\alpha\beta}F^{
\alpha\beta}.\eqno(2.14b)$$
Similarly, for the $W$-exchange bremsstrahlung $bu\to
cd\gamma$, we have
$$\begin{array}{ccl}
{O}^F_\pm(bu\to cd\gamma)=\,{e\over m_i^2-m_f^2} &\Big\{& \left(
e_c{m_f\over m_c}-e_d{m_f\over m_d}\right)\left(\tilde{F}_{\mu\nu}+iF_{\mu\nu}
\right)O_\pm^{\mu\nu}   \\
&+&\left(e_u{m_i\over m_u}-e_b{m_i\over m_b}\right)\left(\tilde{F}_{\mu\nu}
-iF_{\mu\nu}\right)O_\mp^{\mu\nu}\Big\},
\end{array}\eqno(2.15)$$
where now $m_i=m_b+m_u,~m_f=m_c+m_d$.

    Strictly speaking, terms in (2.13) and (2.15) proportional to $1/m_c$ and
$1/m_b$ should have been dropped since we have not included corrections of the
same order to those proportional to $1/m_u$ or $1/m_d$. These $1/m_c$ and
$1/m_b$ terms are kept mainly for phenomenological reasons. We encounter a
similar situation in the electromagnetic decays of heavy hadrons [2,8]
in which the common practice is to retain the $1/m_Q$ contributions due to
the heavy quark's magnetic moments but to neglect the $1/m_Q$ corrections
to the matrix elements of the light quark's electromagnetic currents.
Numerically, the $1/m_c$ terms are especially significant because $m_c$ is
not very large and the $c$ quark carries 2/3 units of charge.
The effective Lagrangians (2.13) and (2.15) are the
main results in this section. We will apply them and (2.1) to the weak
radiative decays of heavy mesons and heavy
baryons in Sections III and IV, respectively.
\vskip 0.4cm
\noindent{\bf III.~~Applications to Heavy Meson Decays}
\vskip 0.4cm

    As shown in the Introduction, the radiative decay modes of interest for
$B$ mesons are $\bar{B}\ri\bar{K}^*\gamma,~\bar{B}_s\to\phi\gamma$ which
receive short-distance contributions from the electromagnetic penguin
$b\ri s\gamma$ transition, and
$\bar{B}\ri D^*\gamma$ proceeding through the $W$
exchange accompanied by a photon emission. The general amplitude of weak
radiative decay with one real photon emission is given by
$$
\begin{array}{ccl}
   A[\bar{B}(p)\ri P^*(q)\gamma(k)] & = & i\epsilon_{\mu\nu\alpha\beta}\es
^\mu k^\nu\es^{*\alpha}q^\beta f_1(k^2)  \\
& + & \es^\mu[\es^*_\mu(m^2_B-m^2_{P^*})-(p+q)_\mu\es^*\cdot k]
f_2(k^2),
\end{array}\eqno(3.1)
$$
where $\es$ and $\es^*$ are the polarization vectors of the photon
and the vector meson $P^*$, respectively, the first (second) term on the r.h.s.
is parity conserving (violating), and $k^2=0$. The decay width implied by
the amplitude (3.1) is
$$\Gamma(\bar{B}\to P^*\gamma)=\,{1\over 32\pi}\,{(m^2_B-m^2_{P^*})^3\over
m^3_B}\,(|f_1|^2+4|f_2|^2).\eqno(3.2)$$
\vskip 0.3cm
\noindent{\bf 3.1~~Effective Lagrangian approach for $\bar{B}^0\to D^{*0}
\gamma$}

    Since the radiative decay $\bar{B}\ri\bar{K}^*\gamma$ has been discussed
extensively in the literature, we will only focus on the second-type mode,
namely $\bar{B}^0\to D^{*0}\gamma$. Our goal is to see if the
tree-level $W$-exchange with a photon emission is
comparable with the short-distance $b\ri s\gamma$ mechanism. We shall use the
factorization method (for a review, see Ref.[9]) to evaluate the hadronic
matrix elements. It follows from Eq.(2.12) that
$$
\begin{array}{ccl}
A(\bar{B}^0\to D^{*0}\gamma) &=& -\la D^{*0}\gamma|{\cal H}_{\rm eff}|\bar{B}
^0\ra  \\      &=& -{G_F\over \sqrt{2}}V_{cb}V_{ud}^*a_2\,{e\over m_i^2-m_f^2}
\la D^{*0}|\bar{c}\gamma_\mu(1-\gamma_5)u|0\ra\la0|\bar{d}\gamma_\nu(1-\gamma
_5)b|\bar{B}^0\ra   \\   &\times & \Bigg[ \tilde{F}^{\mu\nu}\left(e_c{m_f\over
 m_c}+e_d{m_i\over m_d}+e_u{m_f\over m_u}+e_b{m_i\over m_b}\right) \\
&+& iF^{\mu\nu}
\left(e_c{m_f\over m_c}+e_d{m_i\over m_d}-e_u{m_f\over m_u}-e_b{m_i\over m_b}
\right)\Bigg],
\end{array}\eqno(3.3)$$
with
\footnote{In the conventional vacuum insertion method $a_2$ is equal to
$(2c_+-c_-)/3$, while it is $(c_+-c_-)/2$ in the large $N_c$ approximation
in which the Fierz-transformed contributions characterized by the color
factor $1/N_c$ are dropped [9]. The leading $1/N_c$ expansion is known to
work well
for nonleptonic weak decays of charmed mesons. In bottom meson decays, the
magnitude of $a_2$ determined from the measured $\bar{B}\ri\psi\bar{K},~\psi
\bar{K}^*$ rates is in agreement with that predicted by the large $N_c$
approach [10-12]. However, contrary to what expected from the same approach,
the sign of $a_2$ is found to be positive by recent CLEO measurements of
$\bar{B}\ri D\pi,~D\rho,~D^*\pi,
{}~D^*\rho$ decays [11-13]. Thus we take $a_2$ to be that given by (3.4).}
$$a_2=\,{1\over 2}(c_--c_+),\eqno(3.4)$$
and $m_i=m_b+m_d\approx m_B,~m_f=m_c+m_u\approx m_{D^*}$.
The one-body matrix elements appearing in (3.3) have the expressions
$$\la 0|A_\mu|P(p)\ra=\,if_Pp_\mu,\eqno(3.5a)$$
$$\la 0|V_\mu|P^*(p,\es^*)\ra=\,if_Vm_{P^*}\es^*_\mu.\eqno(3.5b)$$
Therefore,
$$\la D^{*0}(p_D)|\bar{c}\gamma_\mu(1-\gamma_5)u|0\ra\la 0|\bar{d}\gamma_\nu(1-
\gamma_5)b|\bar{B}^0(p_B)\ra=-f_Bf_{D^*}m_{_{D^*}}\es_\mu^*p_{_{B\nu}}.
\eqno(3.6)$$
With the substitution of (3.6), (3.3) has the desirable amplitude structure
indicated by (3.1). We find
$$f_1=\,2\eta e\left[ \left({e_u\over m_u}+{e_c\over m_c}\right){m_{D^*}\over
m_B}+\left({e_d\over m_d}+{e_b\over m_b}\right)\right]\,{m_Bm_{D^*}\over
m_B^2-m_{D^*}^2},\eqno(3.7a)$$
$$f_2=\,\eta e\left[ \left({e_u\over m_u}-{e_c\over m_c}\right){m_{D^*}\over
m_B}-\left({e_d\over m_d}-{e_b\over m_b}\right)\right]\,{m_Bm_{D^*}\over
m_B^2-m_{D^*}^2},\eqno(3.7b)$$
with
$$\eta=\,{G_F\over 2\sqrt{2}}V_{cb}V_{ud}^*a_2f_Bf_{D^*}.\eqno(3.8)$$
Substituting (3.7) into (3.2) gives the decay rate
\footnote{The result (3.9) was also obtained by Mendel and Stiarski [14] in
a different approach except for the Wilson factor $a_2$ being replaced
by $(2c_+-c_-)/3$ in the latter work.}
$$
\begin{array}{ccl}
\Gamma(\bar{B}^0\to D^{*0}\gamma) &=& {\alpha\over 32}\,G_F^2f_B^2f_{D^*}^2
|V_{cb}V_{ud}^*|^2m_{D^*}^2m_B\left(1-{m_{D^*}^2\over m_B^2}\right)(c_--c_+)^2
\\    &\times &   \left\{e^2_u{m_{D^*}^2\over m_B^2}\left({1\over m_u^2}+{1
\over m_c^2}\right)+e_d^2\left({1\over m_d^2}+{1\over m_b^2}\right)+2e_ue_d
{m_{D^*}\over m_B}\left({1\over m_um_b}+{1\over m_dm_c}\right)\right\}.
\end{array}\eqno(3.9)$$

   In order to have a numerical estimate, we adopt the following mass
parameters
\footnote{The values of the constituent quark masses are given on p. 1729
of Ref.[15].}
$$
\begin{array}{rcl}
 &m_{B^0}=  5279.0\,{\rm MeV},~~~~~~ m_{D^{*0}}=2006.7\,{\rm MeV}, \\
& m_u=338\,{\rm MeV}, ~~~~m_d=322\,{\rm MeV},~~~~m_s=510\,{\rm MeV},
\end{array}\eqno(3.10)$$
from the Particle Data Group [15], and $m_c=1.6$ GeV, $m_b=5.0$ GeV. Using
$f_{D^*}=200$ MeV, $f_B=190$ MeV, $V_{cb}=0.040$ [16], and
$\tau(B^0)= 1.50\times 10^{-12}s$ [15], we obtain the branching ratio
$${\cal B}(\bar{B}^0\to D^{*0}\gamma)=\,0.92\times 10^{-6}.\eqno(3.11)$$
It is evident that the weak radiative decays of the $\bar{B}$ mesons is
dominated by the electromagnetic penguin diagram. The suppression of $\bar{B}
\to D^*\gamma$ relative to $\bar{B}\to \bar{K}^*\gamma$ is mainly attributed to
the smallness of the decay constants $f_{D^*}$ and $f_B$ occurred in weak
transitions.

    In the factorization approximation the three decay amplitudes $\bar{B}^0
\to D^{*0}\gamma,~\bar{B}^{*0}\to D^{*0}\gamma$ and $\bar{B}^{*0}\to D^0
\gamma$ are all related since $f_P=f_V$ by heavy quark symmetry. However,
the branching ratios for the latter two reactions are expected to be much
smaller than that for $\bar{B}^0\to D^{*0}\gamma$ as $\bar{B}^{*0}$ has a
dominant electromagnetic decay.
\vskip 0.3cm
\noindent{\bf 3.2~~Pole model approach for $\bar{B}^0\to D^{*0}\gamma$}

     Before proceeding further, we would like to compare our present formalism
with the conventional long-distance pole mechanism which has been applied to
the decay $\bar{B}\to \bar{K}^*\gamma$ before [17]. Note that the intermediate
$\bar{B}$ state is absent in Fig. 3d as the $\bar{B}\bar{B}\gamma$ coupling
is prohibited. If we just focus on the
low-lying intermdiate states in the pole diagrams depicted in Fig. 3,
the amplitudes of the first three pole diagrams are
$$
M_a = \la D^{*0}(q)|\lem|D^0(p)\ra\,{1\over m^2_B-m^2_D}\la D^0(p)|\L_W|\bar{
B}^0(p)\ra,\eqno(3.12a)$$
$$M_b = \la D^{*0}(q)|\L_W|\bar{B}^{*0}(q)\ra\,{1\over m^2_{D^*}-m^2_{B^*}}\la
\bar{B}^{*0}(q)|\lem|\bar{B}^0(p)\ra,  \eqno(3.12b)$$
$$M_c = \la D^{*0}(q)|\lem|D^{*0}(p)\ra\,{1\over m^2_B-m^2_{D^*}}\la D^{*0}(p)
|\L_W|\bar{B}^0(p)\ra,  \eqno(3.12c)$$
where the amplitudes $M_a$ and $M_b$ are parity conserving, while $M_c$ is
parity violating.
Since the intermediate pole states are far from their mass shell, we write the
photon couplings in Fig. 3 as
$$
\begin{array}{rcl}
g_{_{DD^*\gamma}}(q^2=m^2_B) &=& g_{_D}(q^2)g_{_{DD^*\gamma}},  \\
g_{_{B^*B\gamma}}(q^2=m^2_D) &=& g_{_{B^*}}(q^2)g_{_{B^*B\gamma}},  \\
g_{_{D^*D^*\gamma}}(q^2=m^2_B) &=& g_{_{D^*}}(q^2)g_{_{D^*D^*\gamma}},
\end{array}\eqno(3.13)$$
where $g_{_{DD^*\gamma}},~g_{_{B^*B\gamma}}$ and $g_{_{D^*D^*\gamma}}$ are
on-shell photon coupling constants which can be calculated in the
nonrelativistic quark model. In Eq.(3.13), $g_{_D},~g_{_{B^*}}$ and $g_{_{D^*
}}$ are form factors accounting for off-shell effects. They are normalized
to unity when mesons are on shell; for example, $g_{_D}(m^2_D)=1,~g_{_{B^*}}
(m^2_{B^*})=1$.

   For $P,P^*=(Q\bar{q})$, we find from Eqs.(2.19) and (2.30) of Ref.[2] that
$$
\la D^{*0}(q)|\lem|D^0(p)\ra = ig_{_D}(m_B^2)g_{_{D^0D^{*0}\gamma}}
\epsilon_{\mu\nu\alpha\beta}\es^\mu k^\nu p^\alpha \es^{*\beta},\eqno(3.14a)$$
$$\la\bar{B}^{*0}(q)|\lem|\bar{B}^0(p)\ra = ig_{_{B^*}}(m_D^2)g_{_{\bar{B}
^{*0}\bar{B}^0\gamma}}\epsilon_{\mu\nu\alpha\beta}\es^\mu k^\nu p^\alpha \es^
{*\beta},\eqno(3.14b)$$
$$\la D^{*0}(q)|\lem|D^{*0}(p)\ra = g_{_{D^*}}(m_B^2)g_{_{D^{*0}D^{*0}
\gamma}}(k_\mu\es_\nu-k_\nu\es_\mu)\es^{*\mu}(q)\es^{*\nu}(p) m_{_{D^*}}.
\eqno(3.14c)$$
Because the intermediate state is far from its mass shell, we have
introduced the form factors defined in Eq.(3.13).
In the heavy quark effective theory, the on-shell $P^*P\gamma$ and $P^*P^*
\gamma$ coupling constants are related to each other and are given by [2]
$$g_{_{PP^*\gamma}}=-2\sqrt{m_{P^*}\over m_P}(e_qd+e_Qd')\equiv e\sqrt{m_{P^*}
\over m_P}(e_q\beta+e_Q\beta'),\eqno(3.15a)$$
$$g_{_{P^*P\gamma}}=-2\sqrt{m_P\over m_{P^*}}(e_qd+e_Qd')\equiv e\sqrt{m_P
\over m_{P^*}}(e_q\beta+e_Q\beta'),\eqno(3.15b)$$
$$g_{_{P^*P^*\gamma}}=-2(e_qd-e_Qd')\equiv e(e_q\beta-e_Q\beta'),
\eqno(3.15c)$$
where $e_q$ is the charge of the light quark $q$ (not $\bar{q}$), and $e_Q$ is
the charge of the heavy quark $Q$. The coupling
$\beta'$ (or $d'$) is fixed by heavy quark symmetry to be $1/m_Q$, while
$\beta$ (or $d$) is independent of heavy quarks and cannot be determined by
heavy quark symmetry alone. In the constituent quark model, $\beta$ is
given by [2]
$$\beta=\,{1\over m_q}.\eqno(3.16)$$
Therefore,
$$g_{_{D^0D^{*0}\gamma}}=\,e\left({e_u\over m_u}+{e_c\over m_c}\right),
\eqno(3.17a)$$
$$g_{_{\bar{B}^{*0}\bar{B}^0\gamma}}=\,e\left({e_d\over m_d}+{e_b\over
m_b}\right),\eqno(3.17b)$$
$$g_{_{D^{*0}D^{*0}\gamma}}=\,e\left({e_u\over m_u}-{e_c\over m_c}
\right),\eqno(3.17c)$$
where the small difference between $m_P$ and $m_{P^*}$ has been neglected.
Note that it is important to keep the contribution from the magnetic moment of
the charmed quark since it is not particularly heavy.
In general, it is expected that the form factors appearing in Eq.(3.14)
become smaller as the hadron is more away from its mass shell owing to less
overlap between initial and final meson wave functions.

  Using the factorization method, we obtain
$$\la D^0(p)|\L_W|\bar{B}^0(p)\ra=\,{G_F\over 2\sqrt{2}}V_{cb}V^*_{ud}(c_--
c_+)f_Df_Bm^2_B,\eqno(3.18)$$
and use has been made of $p^2=m^2_B$. Likewise, we obtain
$$\la D^{*0}(q)|\L_W|\bar{B}^{*0}(q)\ra=\,{G_F\over 2\sqrt{2}}V_{cb}V^*_{ud}
(c_--c_+)f_Df_Bm_Dm_B(\es^*_B\cdot\es^*_D),\eqno(3.19a)$$
$$\la D^{*0}(p)|\L_W|\bar{B}^{0}(p)\ra=\,-{G_F\over 2\sqrt{2}}V_{cb}V^*_{ud}
(c_--c_+)f_Df_Bm_{D^*}(p\cdot\es^*_D).\eqno(3.19b)$$
Putting everything together and using the relation
$$\sum_\lambda \es^*_\mu(q)\es^*_\nu(q)=-g_{\mu\nu}+{q_\mu q_\nu\over
m^2},\eqno(3.20)$$
we finally find
$$f_1(0) =2\eta\,{f_{_D}\over f_{_{D^*}}}\left[{m^2_B\over m_B^2-m_D^2}g_{_{
D^0D^{*0}\gamma}}g_{_{D}}(m^
2_B)+{m_Dm_B\over m_{B^*}^2-m^2_{D^*}}g_{_{\bar{B}^{*0}\bar{B}^0\gamma}}g_{_{
B^*}}(m^2_D)\right],\eqno(3.21a)$$
$$f_2(0) = -\eta\,{f_{_D}\over f_{_{D^*}}}g_{_{D^{*0}D^{*0}\gamma}}g_{_{D^*}}
(m^2_B).\eqno(3.21b)$$
Comparing (3.21) with (3.7) yields
$$g_{_D}(m^2_B)={m_D^2\over m_B^2},~~~g_{_{B^*}}(m_D^2)=1,~~~g_{_{D^*}}(m_B^2)
=-{m_{D^*}^2\over m_B^2-m_{D^*}^2},\eqno(3.22)$$
where use of the heavy quark symmetry relations $m_{B^*}=m_B$, $m_{D^*}=
m_D$ and $f_{D^*}=f_D$
has been made. We note that in the effective Lagrangian approach there is
an additional term proportional to $(e_d/m_d-e_b/m_b)$ in $f_2$. What is
the counterpart of this term in the pole model? Evidently, it must come
from a $p$-wave $1^+$ $\bar{B}$ resonance state (see Fig. 3d): The $E$1
$\bar{B}^0(1^+)\bar{B}^0(0^-)\gamma$ transition coupling is proportional to
$(e_d/m_d-e_b/m_b)$ provided that $\bar{B}(1^+)$ is a $p$-wave spin-singlet,
while $\bar{B}(1^+)-D^*$ weak transition is parity violating.
We thus see that both approaches are consistent with
each other. However, the effective Lagrangian approach is simpler and it
also provides the information on the form factors, as shown in (3.22).

\vskip 0.4cm
\noindent{\bf IV. ~~Applications to Heavy Baryon Decays}
\vskip 0.4cm
In this section we will first focus our attention on the short-distance
penguin effect
 in the decays $\Lambda_b\ri\Sigma\gamma$ and $\Lambda_b\ri\Lambda\gamma$ and
then turn to the weak radiative decays of the antitriplet bottom baryons,
namely $\Xi_b^0\to \Xi_c^0({\Xi'}_c^0)\gamma$ and $\Lambda_b^0\to\Sigma_c^0
\gamma$. Since the weak radiative decay of $\bar{B}$ mesons is dominated by
the short-distance
$b\to s\gamma$ transition, it is natural to expect that the same mechanism
also works for bottom baryons. The general amplitude of baryon weak
radiative decay reads
$$A(B_i\to B_f\gamma)=\,i\bar{u}_f(a+b\gamma_5)\sigma_{\mu\nu}\es^\mu k^
\nu u_i,\eqno(4.1)$$
where $a$ and $b$ are parity-conserving and -violating amplitudes,
respectively. The corresponding decay rate is
$$\Gamma(B_i\to B_f\gamma)=\,{1\over 8\pi}\left( {m^2_i-m^2_f\over m_i}\right)
^3(|a|^2+|b|^2).\eqno(4.2)$$
\vskip 0.2cm
\noindent{\bf 4.1~~Penguin induced baryon radiative decays}

   The electromagnetic penguin transition $b\to s+\gamma$ at the quark
level is well understood in QCD, and its effective Lagrangian is given by
(2.1). However, its hadronic matrix elements depend on the nature of
the bound states, and there are no known methods to calculate them from first
principles. In this Section, we will use two different methods to estimate
the rates for the decay
$\Lambda_b\to\Lambda+\gamma$. In the first method, both the $b$ and $s$
quarks are treated as heavy. Heavy quark symmetry then gives model
independent prediction. Although we are able to compute the correction of
order $m_s/m_b$ and $1/m_s$, higher order corrections are needed for a
realistic comparsion with experiment. They are in principle well defined in
QCD; but it is impossible to compute them all. In the second method, only the
$b$ quark is treated as heavy so the form factors of tensor currents needed
are related by heavy quark symmetry to those of vector and axial vector
currents, which are then evaluated in the MIT bag model. The MIT bag model
includes the $m_s$ effects to all orders, but the reliability of the
model is hard to assess. It is clear that neither of the two
approaches is very satisfactory. Together, however, they provide a more
or less consistent order of magnitude estimate of these weak radiative
decays.

   In the first method we treat the $s$ quark as a heavy quark and then
take into account the $1/m_s$ and QCD corrections. Despite that the
effective mass of the $s$ quark in hyperons is only of order 500 MeV, it is
not small compared to the QCD scale and we thus expect to see some vestiges
of heavy quark symmetry. In the heavy $s$ quark
limit, the hyperon $\Lambda$ behaves as an antitriplet heavy baryon
$B_{\bar{3}}$, while $\Sigma^0$ as a sextet baryon $B_6$.
{}From Eq.(2.1) we obtain
$$\begin{array}{rcl}
A(\Lambda^0_b\to {\rm hyperon}+\gamma)_{\rm SD} &=& i{G_F\over \sqrt{2}}\,{e
\over 8\pi^2}F_2V_{tb}V_{ts}^*\,m_b\es^\mu k^\nu \\
&\times& \la{\rm hyperon}|\bar{s}\sigma_{\mu\nu}[(1+\gamma_5)+{m_s\over m_b}
(1-\gamma_5)]b|\Lambda_b^0\ra.
\end{array}\eqno(4.3)$$
Using the interpolating fields [18]
$$B_{\bar{3}}(v,s)=\,\bar{u}(v,s)\phi_vh_v,\eqno(4.4a)$$
$$B_6(v,s)=\,\bar{B}_\mu(v,s)\phi^\mu_v h_v,\eqno(4.4b)$$
where $\phi_v$ and $\phi^\mu_v$ are the $0^+$ and $1^+$ diquarks, respectively,
which combine with the heavy quark $h_v$ of velocity $v$ to form the
appropriate heavy baryon, and the relations [18]
$$\la 0|\phi_{v'}\phi^\dagger_v|0\ra =\, \zeta(v\cdot v'),\eqno(4.5a)$$
$$\la 0|\phi_{v'}^\mu\phi_v^\dagger|0\ra =\, 0,\eqno(4.5b)$$
where $\zeta(v\cdot v')$ is a universal baryonic Isgur-Wise function
normalized to unity at the zero recoil $v\cdot v'=1$, we find
\footnote{In the heavy $s$ quark limit, a $\Lambda$ is made of just a strange
quark and a scalar diquark with $(ud)$ quantum numbers. Contrary to some
claims made in the literature, it is not necessary to include a Clebsch-Gordon
coefficient $1/\sqrt{3}$ in Eq.(4.6a).}
$$\la\Lambda(v',s')|\bar{s}\sigma_{\mu\nu}(1\pm\gamma_5)b|\Lambda^0_b(v,s)\ra
=\,\bar{u}_{_\Lambda}\sigma_{\mu\nu}(1\pm\gamma_5)u_{_{\Lambda_b}}\zeta(v\cdot
v'),\eqno(4.6a)$$
$$\la\Sigma^0(v',s')|\bar{s}\sigma_{\mu\nu}(1\pm\gamma_5)b|\Lambda^0_b(v,s)\ra
=\,0.\eqno(4.6b)$$
Therefore, no weak $B_{\bar{3}}-B_6$ transition can be induced by the $b\ri
s\gamma$ mechanism in the heavy quark limit. So, the first prediction we have
is
$$\Gamma(\Lambda_b^0\to\Sigma^0\gamma)<<\Gamma(\Lambda_b^0\to\Lambda^0\gamma).
\eqno(4.7)$$

   We will follow Ref.[19] to treat the $1/m_s$ corrections to $\Lambda_b^0\to
\Lambda^0\gamma$. First, from the
relationship between the $s$ quark field and the effective field $h_{v'}^{
(s)}$
$$s(x)=\,e^{-im_sv'\cdot x}\left[ 1+{1-v\!\!\!/'\over 2}\,{iD\!\!\!\!/ \over
2m_s}\right]h_{v'}^{(s)},\eqno(4.8)$$
we get
$$\bar{s}\sigma_{\mu\nu}(1\pm\gamma_5)b \to \bar{h}^{(s)}_{v'}\left(1-{i\D
\over 2m_s}\right)\sigma_{\mu\nu}(1\pm\gamma_5)h^{(b)}_v.\eqno(4.9)$$
Applying the result [19]
$$\la\Lambda(v',s')|\bar{h}^{(s)}_{v'}i\D \Gamma h^{(b)}_v|\Lambda_b^0(v,s)\ra
=\bar{\Lambda}\zeta(v\cdot v')\,{v_\mu-(v\cdot v')v'_\mu\over 1+v\cdot v'}\,
\bar{u}_{_\Lambda}(v',s')\gamma^\mu\Gamma u_{_{\Lambda_b}}(v,s),\eqno(4.10)$$
with the new parameter $\bar{\Lambda}$ being
$$\bar{\Lambda}=\,m_{_{\Lambda_b}}-m_b=\,m_{_{\Lambda_c}}-m_c=\,m_{\Lambda}
-m_s\approx 700~{\rm MeV},\eqno(4.11)$$
we obtain
$$
\begin{array}{rcl}
&&\la\Lambda(v',s')|\bar{h}^{(s)}_{v'}i\D \sigma_{\mu\nu}\es^\mu k^\nu
(1\pm\gamma_5) h^{(b)}_v|\Lambda_b^0(v,s)\ra \\ &=& h\bar{\Lambda}\zeta(v\cdot
v')\,\bar{u}_{_\Lambda}(v',s')\sigma_{\mu\nu}\es^\mu k^\nu(1\pm\gamma_5)u_{_{
\Lambda_b}}(v,s),\end{array}\eqno(4.12)$$
with $h=({m_{\Lambda}\over m_{\Lambda_b}}-v\cdot v')/(1+v\cdot v')$,
where use has been made of
$$v=\,{m_\Lambda\over m_{\Lambda_b}}v'+{1\over m_{\Lambda_b}}k,~~~~~~\es\!\!\!/
k\!\!\!/=-k\!\!\!/\es\!\!\!/.\eqno(4.13)$$
It follows from Eqs.(4.3), (4.6), (4.9) and (4.12) that
$$ a=\, {G_F\over \sqrt{2}}\,{e\over 8\pi^2}F_2m_bV_{tb}V^*_{ts}\left(1+{m_s
\over m_b}-{\bar{\Lambda}\over 2m_s}h\right)
\zeta(v\cdot v'),\eqno(4.14a)$$
$$   b=\, {G_F\over \sqrt{2}}\,{e\over 8\pi^2}F_2m_bV_{tb}
V^*_{ts}\left(1-{m_s\over m_b}-{\bar{\Lambda}\over 2m_s}h\right)\zeta(v\cdot
v').\eqno(4.14b)$$
Including QCD corrections gives rise to [20]
$$\zeta(v\cdot v')=\,C(\mu)\zeta_0(v\cdot v',\mu),\eqno(4.15)$$
where
$$C(\mu) = \left({\alpha_s(m_b)\over \alpha_s(m_c)}\right)^{-6/25}
\left({\alpha_s(m_c)\over \alpha_s(m_s)}\right)^{-6/27}
\left({\alpha_s(m_s)\over \alpha_s(\mu)}\right)^{a_L},\eqno(4.16a)$$
$$a_L(\omega) = {8\over 29}\left({\omega\over\sqrt{\omega^2-1}}\ln(\omega+
\sqrt{\omega^2-1})-1\right),\eqno(4.16b)$$
with $\omega\equiv v\cdot v'$.
There is no obvious choice for the normalization scale in Eq.(4.15). It is
expected that there will be no large parameters in the function $\zeta_0(
\omega,\mu)$ if the renormalization scale is low [19]. However, since
perturbation theory will break down at very low scales, we thus choose $\mu
\sim m_s$ so that $\alpha_s(\mu)\sim 1$, and $C(\mu)\simeq 1.23\,$.
For the decay $\Lambda_b\to\Lambda\gamma$, $\omega=2.63\,$.  It is easily seen
that the $1/m_s$ correction to the $\Lambda_b\to\Lambda\gamma$ amplitude is
about $50\%$ for $m_s=510$ MeV, which is quite sizeable. This implies that
it is important to include higher order $1/m_s$ corrections. However, this is
beyond the scope of the present paper.

 In order to estimate the decay rate of $\Lambda_b\to\Lambda\gamma$, we employ
two recent models for $\zeta(v\cdot v')$:
$$\zeta_0(\omega) =\,0.99\exp[-1.3(\omega-1)]~~~~{\rm (soliton~model~[21])},
\eqno(4.17a)$$
$$\zeta_0(\omega) =\,\left( {2\over \omega+1}\right)^{3.5+{1.2\over\omega}}~~~~
{\rm (MIT~bag~model~[22])}.\eqno(4.17b)$$
Hence, $\zeta_0(\omega=2.63)$ ranges from 0.09 to 0.12 . Substituting (4.14)
into (4.2) and using the good approximation $V_{tb}V_{ts}^*\cong -V_{cb}V^*_{
cs}$ and the lifetime $\tau(\Lambda_b)=\,1.07\times 10^{-12}\,s$ [15], we find
$${\cal B}(\Lambda^0_b\to\Lambda^0\gamma)=\,1.31\times 10^{-3}|\zeta_0(\omega)|
^2=\,(1.2-1.9)\times 10^{-5}.\eqno(4.18)$$
Since there is only one strange quark in $\Xi_b$ but two strange quarks in
$\Omega_b$ and $\Xi$, it is not clear to us how to generalize the above heavy
$s$ quark method to the radiative decays $\Xi_b\to\Xi\gamma$ and $\Omega_b\to
\Omega\gamma$.

     In the second method, only the $b$ quark is treated as heavy. Since
$\gamma_0b=b$ in the static limit of the $b$ quark, we have the relation
$$
\la\Lambda|\bar{s}i\sigma_{0i}(1\pm\gamma_5)b|\Lambda_b\ra=\,\la\Lambda|\bar{s}
\gamma_i(1\mp\gamma_5)b|\Lambda_b\ra.\eqno(4.19)$$
In the heavy $b$-quark limit, there are only two independent form factors
in the $\gamma_\mu$ and $\gamma_\mu\gamma_5$ matrix elements [23]:
$$\la\Lambda(p)|\bar{s}\gamma_\mu(1\mp\gamma_5)b|\Lambda_b(v)\ra=\,\bar{u}
_{_\Lambda}\left(F_1^{\Lambda_b\Lambda}(v\cdot p)+v\!\!\!/ F_2^{\Lambda_b
\Lambda}(v\cdot p)\right)\gamma_\mu(1\mp\gamma_5)u_{_{\Lambda_b}}.\eqno(4.20)$$
Recall that $F_1=1$ at zero recoil and $F_2=0$ if both $\Lambda_b$ and
$\Lambda$ are treated as heavy. The relationship between $F_{1,2}$ and
the standard form factors defined by
$$\begin{array}{ccl}
\la\Lambda(p)|\bar{s}\gamma_\mu(1-\gamma_5)b|\Lambda_b(p')\ra &=& \bar{u}
_{_\Lambda}\big[f_1(q^2)\gamma_\mu+if_2(q^2)\sigma_{\mu\nu}q^\nu+
f_3(q^2)q_\mu \\    && -(g_1(q^2)\gamma_\mu+ig_2(q^2)\sigma_{\mu\nu}q^\nu
+g_3(q^2)q_\mu)\gamma_5\big]u_{_{\Lambda_b}},\end{array}\eqno(4.21)$$
reads
$$\begin{array}{ccl}
&& f_1(q^2)=g_1(q^2)=F_1(q^2)+{m_\Lambda\over m_{\Lambda_b}}F_2(q^2), \\
&& f_2(q^2)=g_2(q^2)=f_3(q^2)=g_3(q^2)=\,{1\over m_{\Lambda_b}}F_2(q^2).
\end{array}\eqno(4.22)$$
It follows from (4.3), (4.19) and (4.20) that
$$ a=\, {G_F\over \sqrt{2}}\,{e\over 8\pi^2}F_2m_bV_{tb}V^*_{ts}\left(1+{m_s
\over m_b}\right)\left[F_1^{\Lambda_b\Lambda}(0)-F_2^{\Lambda_b\Lambda}(0)
\right],\eqno(4.23a)$$
$$   b=\, {G_F\over \sqrt{2}}\,{e\over 8\pi^2}F_2m_bV_{tb}
V^*_{ts}\left(1-{m_s\over m_b}\right)\left[F_1^{\Lambda_b\Lambda}(0)-F_2^{
\Lambda_b\Lambda}(0)\right],\eqno(4.23b)$$
where $F(0)$ here means $F(q^2=0)$.

   In principle, the Isgur-Wise form factors $F_{1,2}(0)$ can be obtained by
first computing the form factors $f_i$ and $g_i$ at zero recoil in a quark
model
\footnote{This is because the quark model calculations are presumably most
reliable at zero recoil where both baryons are static.}
and then extrapolating to $q^2=0$ under some assumption on their $q^2$
dependence. In the literature, form factors $f_i$ and $g_i$ have been
calculated for $\Lambda_c\to\Lambda$ transition in two different quark models:
the nonrelativistic quark model and the MIT bag model [24]. The model
calculation can be tested by a recent CLEO measurement [25]. Assuming a dipole
$q^2$ dependence of the form factors
$$F_i(q^2)=\,{F_i(0)\over (1-q^2/m_*^2)^2},\eqno(4.24)$$
where $m_*$ is a pole mass,
\footnote{Beyond the heavy quark limit, the pole masses $m_V$ and $m_A$ for
form factors $f_i$ and $g_i$ respectively are different. In general, the pole
mass is taken to be the mass of the nearest spin-one meson with the right
quantum number. For $\Lambda_c\to\Lambda$ transition, $m_V=m_{D_s(1^-)}=2.11$
GeV, $m_A=m_{D_s(1^+)}=2.536$ GeV, while for $\Lambda_b\to\Lambda$: $m_V=
m_{B_s(1^-)}\cong 5.42$ GeV, $m_A=m_{B_s(1^+)}\cong 5.86$ GeV.}
the form factor ratio $R\equiv F_2^{\Lambda_c\Lambda}/
F_1^{\Lambda_c\Lambda}$ is found to be
$$R=-0.33\pm 0.16\pm 0.15\eqno(4.25)$$
by CLEO for $m_*=2.11$ GeV [25]. From Tables IV and VI of Ref.[24], we find
\footnote{Note that the definition of the form factors $f_i$ and $g_i$ in
Ref.[24] follows that in Ref.[26] and is different from ours. In terms of our
notation, the bag model predictions shown in Table VI for $\Lambda_c\to
\Lambda$ form factors evaluated at $q^2=0$ with dipole approximation are
$f_1=0.46,~f_2=-0.19/m_{\Lambda_c},~g_1=0.50,~g_2=-0.05/m_{\Lambda_c}$,
while the nonrelativistic quark model predicts $f_1=0.35,~f_2=
-0.09/m_{\Lambda_c},~g_1=0.61,~g_2=-0.04/m_{\Lambda_c}$. The difference
between $f_i$ and $g_i$ is attributed to $1/m_c$ corrections. The form factor
ratio calculated in (4.26) comes from the form factors $f_1$ and $f_2$.}
$$R=\cases{-0.34\,, &MIT~bag~model; \cr -0.23\,, &nonrelativistic~quark model.
\cr}\eqno(4.26)$$
It is clear that while both models' predictions are consistent with
experiment, the bag model gives a better agreement.

    As for the $\Lambda_b\to\Lambda$ form factors, our strategy is as follows.
We will use the bag model to compute $f_1^{\Lambda_b\Lambda}$ at zero recoil,
and then employ the heavy-flavor-symmetry relation
$$F_2^{\Lambda_c\Lambda}(q^2_{\Lambda_c})/F_1^{\Lambda_c\Lambda}(q^2_{
\Lambda_c})=\,F_2^{\Lambda_b\Lambda}(q^2_{\Lambda_b})/F_1^{\Lambda_b\Lambda}
(q^2_{\Lambda_b}),\eqno(4.27)$$
where $q_{\Lambda_c}=m_{\Lambda_c}v-q$ and $q_{\Lambda_b}=m_{\Lambda_b}v-q$,
together with the CLEO data for $R$, to determine $F_{1,2}^{\Lambda_b\Lambda}$
at $q^2_m\equiv (m_{\Lambda_b}-m_\Lambda)^2$, which can be extrapolated to
$q^2=0$ using Eq.(4.24). In the MIT bag model [27], $f_1$ at zero recoil is
given by
\footnote{A small and negligible correction to (4.28) is shown in Eq.(5a) of
Ref.[24].}
$$f_1^{\Lambda_b\Lambda}(q^2_m)=\int d^3r\left(
u_b(r)u_s(r)+v_b(r)v_s(r)\right),\eqno(4.28)$$
where $u(r),~v(r)$ are respectively the large and small components of the
$1S_{1/2}$ quark spatial wave function.
Numerically, we find $f_1^{\Lambda_b\Lambda}(q^2_m)=0.95$ (see e.g. Ref.[28]
for the technique). It follows from (4.22), (4.25) and (4.27) that
$$F_1^{\Lambda_b\Lambda}(q^2_m)=1.02\,,~~~~~F_2^{\Lambda_b\Lambda}(q^2_m)=
-0.34\,,\eqno(4.29)$$
where only the central values are presented. Since the form factor ratio is
measured by CLEO using the dipole $q^2$ dependence, it is
natural to employ the same dipole behavior for $F_i^{\Lambda_b\Lambda}$ for
the sake of consistency. Note that the dipole form
$$G(q^2)={\left(1-q_m^2/m^2_*\right)^2\over \left(1-
q^2/m_*^2\right)^2}\eqno(4.30)$$
plays the role of the baryonic Isgur-Wise function
$\zeta(v\cdot v')$. Using $m_*=5.42$ GeV, we find $G(0)=0.092$, in accordance
with $\zeta(2.63)=0.09\sim 0.12$. Putting everything together,
we finally obtain
$${\cal B}(\Lambda_b\to\Lambda+\gamma)=\,0.6\times 10^{-5}.\eqno(4.31)$$
This model result is close to the previous prediction (4.18).

   We should accentuate the nature of the above two calculations. The first
method is model independent; but since the effective $s$ quark in the
baryon is only of order 500 MeV, it is questionable to apply the heavy quark
effective theory to hyperons, as evidenced by the sizeable $1/m_s$
corrections shown above. The second method takes care of the $m_s$ effects
to all orders, but it is hard to assess its reliability. For example,
quark model predictions for the form factors $f_i$ and $g_i$ depend on
the model content; the form-factor $q^2$ dependence is very sensitive to the
choice of the pole behavior: monopole or dipole. Nevertheless, we see that
both approaches provide a consistent order of magnitude estimate; that is,
${\cal B}(\Lambda_b\to\Lambda\gamma)\sim 1\times 10^{-5}$.

\vskip 0.3cm
\noindent{\bf 4.2~~Heavy quark symmetry predictions for $\Xi_b^0\to\Xi_c^0
\gamma$}

     Before embarking on quark model calculations for the radiative decays
$\Xi_b^0\to\Xi_c^0\gamma,~{\Xi'}_c^0\gamma$ and $\Lambda_b^0\to\Sigma_c^0
\gamma$,
we would like to see what we can learn from applying the heavy quark symmetry
to these decays. It turns out that for the antitriplet to antitriplet
radiative transition $\Xi_b^0\to\Xi_c^0\gamma$,
heavy quark symmetry implies a nontrivial model independent prediction for
$a/b$, the ratio of the parity-conserving and parity-violating amplitudes.
\footnote{We have checked explicitly that heavy quark symmetry alone does not
lead to any useful predictions for other decays such as $\Xi_b^0\to{\Xi'}_c^0
\gamma,~\Lambda_b^0\to\Sigma_c^0\gamma$.}

  Let us denote
$$O_{1\mu\nu}=\,\bar{c}\gamma_\mu(1-\gamma_5)b\,\bar{d}\gamma_\nu(1-\gamma_5)
u,\eqno(4.32a)$$
$$O_{2\mu\nu}=\,\bar{c}\gamma_\mu(1-\gamma_5)u\,\bar{d}\gamma_\nu(1-\gamma_5)
b,\eqno(4.32b)$$
and apply the interpolating field (4.4a) to the antitriplet heavy baryons
to get
$$\begin{array}{ccl}
\la\Xi_c^0(v')|O_{1\mu\nu}|\Xi_b^0(v)\ra &=& \la 0|\bar{u}_f(v',s')\phi_{v'}
c_{v'}\bar{c}_{v'}\gamma_\mu(1-\gamma_5)b_v\bar{d}\gamma_\nu(1-\gamma_5)u\bar
{b}_v\phi^\dagger_vu_i(v,s)|0\ra  \\
&=& \bar{u}_f(v',s'){1+v\!\!\!/'\over 2}\gamma_\mu(1-\gamma_5){1+v\!\!\!/\over
2}u_i(v,s)\la 0|\phi_{v'}\bar{d}\gamma_\nu(1-\gamma_5)u\phi^\dagger_v|0\ra.
\end{array}\eqno(4.33)$$
Lorentz invariance implies that
$$\la 0|\phi_{v'}\bar{d}\gamma_\nu(1-\gamma_5)u\phi^\dagger_v|0\ra=\,A(v\cdot
v')v_\nu+B(v\cdot v')v'_\nu.\eqno(4.34)$$
Therefore,
$$\la\Xi_c^0(v')|O_{1\mu\nu}|\Xi_b^0(v)\ra =\,\bar{u}_f(v',s')\gamma_\mu(1-
\gamma_5)u_i(v,s)[A(v\cdot v')v_\nu+B(v\cdot v')v'_\nu].\eqno(4.35)$$
Likewise, the matrix element of $O_{2\mu\nu}$ is
$$\la\Xi_c^0(v')|O_{2\mu\nu}|\Xi_b^0(v)\ra =\,\bar{u}_f(v',s')\gamma_\mu(1-
\gamma_5)\la 0|\phi_{v'}u\bar{d}\phi^\dagger_v|0\ra\gamma_\nu(1-\gamma_5)
u_i(v,s).\eqno(4.36)$$
Again, Lorentz invariance demands that
$$\la 0|\phi_{v'}u\bar{d}\phi^\dagger_v|0\ra= \,A'(v\cdot v')v\!\!\!/+B'(v\cdot
v')v\!\!\!/'+C'(v\cdot v')v\!\!\!/' v\!\!\!/+D(v\cdot v').\eqno(4.37)$$

   Our next task is to recast (4.35) and (4.36) into a more suitable form.
Since $O_{1\mu\nu}$ and $O_{2\mu\nu}$ are multiplied by $F^{\mu\nu}$ and
$\tilde{F}^{\mu\nu}$ [see (2.15)], only the antisymmetric part will contribute.
Thus we write
$$\la\Xi_c^0(v')|O_{1\mu\nu}|\Xi_b^0(v)\ra ={1\over 2}\bar{u}_f(v',s')[A(\gamma
_\mu v_\nu-\gamma_\nu v_\mu)+B(\gamma_\mu v'_\nu-\gamma_\nu v'_\mu)](1-
\gamma_5)u_i(v,s).\eqno(4.38)$$
By virtue of the equation of motion $v\!\!\!/ u(v,s)=u(v,s)$, two useful
relations can be derived:
$$(\gamma_\mu v_\nu-\gamma_\nu v_\mu)(1-\gamma_5)=-{i\over 2}\sigma_{\mu\nu}
(1+\gamma_5)+{i\over 2}v\!\!\!/ \sigma_{\mu\nu}(1-\gamma_5),\eqno(4.39a)$$
$$(\gamma_\mu v'_\nu-\gamma_\nu v'_\mu)(1-\gamma_5)={i\over 2}\sigma_{\mu\nu}
(1-\gamma_5)-{i\over 2}\sigma_{\mu\nu}(1+\gamma_5)v\!\!\!/',\eqno(4.39b)$$
which can be further simplified by applying the analog of Eq.(4.13) to the
$\Xi_b\to\Xi_c$ transition. As a consequence of (4.33) and (4.39), we obtain
$$\la\Xi_c^0(v')|O_{1\mu\nu}|\Xi_b^0(v)\ra =-{i\over 4}\left(A+B\mm\right)
\bar{u}_f(v',s')[\sigma_{\mu\nu}(1+\gamma_5)-\m\sigma_{\mu\nu}(1-\gamma_5)]
u_i(v,s).\eqno(4.40)$$

   As for the matrix element of $O_{2\mu\nu}$, (4.36) leads to
$$\begin{array}{ccl}
\la\Xi_c^0(v')|O_{2\mu\nu}|\Xi_b^0(v)\ra  &=&\bar{u}_f(v',s')\left[ A'(\gamma
_\mu v\!\!\!/\gamma_\nu-\gamma_\nu v\!\!\!/\gamma_\mu)+B'(\gamma_\mu
v\!\!\!/'\gamma_\nu-\gamma_\nu v\!\!\!/'\gamma_\mu)\right](1-\gamma_5)u_i(v,s)
 \\
&=& i\bar{u}_f(v',s')\Big\{ A'[(1+v\!\!\!/)\sigma_{\mu\nu}+(1-v\!\!\!/)
\sigma_{\mu\nu}\gamma_5)]  \\      &+& B'\sigma_{\mu\nu}[(1+v\!\!\!/')-\gamma_5
(1-v\!\!\!/')]\Big\}u_i(v,s).
\end{array}\eqno(4.41)$$
Applying the analog of (4.13) again to (4.41) leads to
$$\la\Xi_c^0(v')|O_{2\mu\nu}|\Xi_b^0(v)\ra =\,i\left(A'+\mm B'\right)\bar{u}_f
(v',s')\sigma_{\mu\nu}\left[(1+\gamma_5)+\m(1-\gamma_5)\right]u_i(v,s).
\eqno(4.42)$$
    Finally, substituting (4.40) and (4.42) into (2.15), we obtain
$$\begin{array}{ccl}
\la\Xi_c^0(v')|O^F_\pm|\Xi_b^0(v)\ra &=& {2e\over m_i^2-m_f^2}\left[{1\over
4}\left(A+B\mm\right)\mp\left(A'+B'\mm\right)\right] \\  &\times &
\bar{u}_f(v',s')\sigma\cdot F
\Bigg\{ \left[e_c{m_f\over m_c}-e_d{m_f\over m_d}+\left(e_u
{m_i\over m_u}-e_b{m_i\over m_b}\right)\m\right]   \\
&+& \left[e_c{m_f\over m_c}-e_d{m_f\over m_d}-\left(e_u{m_i\over m_u}-e_b{
m_i\over m_b}\right)\m\right]\gamma_5\Bigg\}u_i(v,s).
\end{array}\eqno(4.43)$$
{}From (4.43) we see that although the $\bar{3}\to\bar{3}+\gamma$ transition
depends on the unknown parameters $A,~B,~A'$ and $B'$, a unique tree-level
prediction on the ratio of $a/b$ based on heavy quark symmetry is nevertheless
accomplished:
$$\left.{a\over b}\right|_{\Xi_b^0\to\Xi_c^0\gamma}=\,{e_c{m_f\over m_c}-e_d
{m_f\over m_d}
+\left(e_u{m_i\over m_u}-e_b{m_i\over m_b}\right)\m\over e_c{m_f\over m_c}-e_d
{m_f\over m_d}-\left(e_u{m_i\over m_u}-e_b{m_i\over m_b}\right)\m}.
\eqno(4.44)$$
This ratio can be tested by measuring the
asymmetry parameter $\alpha$ as will be discussed later.
\vskip 0.3 cm
\noindent{\bf 4.3~~Bag model calculations for $\Xi_b^0\to\Xi_c^0({\Xi'}_c^0)
\gamma$ and $\Lambda_b^0\to\Sigma_c^0\gamma$}

  Recall from Sect. II that the effective Hamiltonian responsible for the weak
radiative decays of heavy baryons is given by
$${\cal H}_{\rm eff}(bu\to cd\gamma)=\,{G_F\over 2\sqrt{2}}V_{cb}V_{ud}^*(
c_+O^F_++c_-O^F_-),\eqno(4.45)$$
where [cf. Eq.(2.15)]
$$O^F_\pm=\,d_f(\tilde{F}_{\mu\nu}+iF_{\mu\nu})O^{\mu\nu}_\pm+d_i(\tilde{F}_{
\mu\nu}-iF_{\mu\nu})O^{\mu\nu}_\mp,\eqno(4.46)$$
with $O^{\mu\nu}_\pm=O_1^{\mu\nu}\pm O_2^{\mu\nu},$ and
$$d_i=\,m_i\left( {e_u\over m_u}-{e_b\over m_b}\right){e\over m_i^2-m_f^2},
\eqno(4.47a)$$
$$d_f=\,m_f\left( {e_c\over m_c}-{e_d\over m_d}\right){e\over m_i^2-m_f^2},
\eqno(4.47b)$$
and $m_i=m_b+m_u,~m_f=m_c+m_d$.
Writing
$$\la B_f(v')|O^{\mu\nu}_{1,2}|B_i(v)\ra=\,i\bar{u}_f(v',s')(a_{1,2}+b_{1,2}
\gamma_5)\sigma^{\mu\nu}u_i(v,s),\eqno(4.48)$$
we get
$$\begin{array}{ccl}
\la B_f(v')|O^F_\pm|B_i(v)\ra &=& -\bar{u}_f(v',s')\Big\{d_f(a_\pm+b_\pm)-d_i
(a_\mp-b_\mp)    \\     && +\left[d_f(a_\pm+b_\pm)+d_i(a_\mp-b_\mp)\right]
\gamma_5\Big\}\sigma\cdot F u_i(v,s),
\end{array}\eqno(4.49)$$
where $a_\pm=a_1\pm a_2,~b_\pm=b_1\pm b_2$, corresponding to the
parity-conserving and parity-violating matrix elements of $O_\pm^{\mu\nu}$.
It follows from Eqs.(4.45) and (4.1) that
$$
a = -{G_F\over \sqrt{2}}V_{cb}V_{ud}^*\left\{ c_+[d_f(a_++b_+)-d_i(a_--b_-)]
+c_-[d_f(a_-+b_-)-d_i(a_+-b_+)]\right\},\eqno(4.50a)$$
$$b= -{G_F\over \sqrt{2}}V_{cb}V_{ud}^*\left\{ c_+[d_f(a_++b_+)+d_i(a_--b_-)]
+c_-[d_f(a_-+b_-)+d_i(a_+-b_+)]\right\}.\eqno(4.50b)$$

  We shall employ the MIT bag model [27] to evaluate the four baryon matrix
elements
$a_\pm$ and $b_\pm$. Since the quark-model wave functions best resemble
the hadronic states in the frame where both baryons are static, we thus adopt
the static bag approximation for the calculation. Because
$$\bar{u}_f^\up\sigma^{xy} u_i^\up=1,~~~~\bar{u}_f^\up\sigma^{0z}\gamma_5
u_i^\up=i,\eqno(4.51)$$
for $\vec{v}_f=\vec{v}_i=0$, it follows from (4.48) that
$$a_{1,2} = -i\la B_f\up|O_{1,2}^{xy}|B_i\up\ra,\eqno(4.52a)$$
$$b_{1,2} = -\la B_f\up|O_{1,2}^{0z}|B_i\up\ra.\eqno(4.52b)$$

   Matrix elements $a_{1,2}$ and $b_{1,2}$ in the MIT bag model can be
expressed in terms of four-quark overlap bag
integrals (see e.g. Ref.[28] for the method).
To do this we need the spin-color wave functions
of the baryons involved such as
$$\begin{array}{ccl}
\Xi_b^0 &=& {1\over\sqrt{6}}[(bus-bsu)\chi_A+(12)+(13)], \\
\Sigma_c^0 &=& {1\over\sqrt{3}}[cdd\chi_s+(12)+(13)],
\end{array}\eqno(4.53)$$
where $abc\chi_s=(2a^\dw b^\up c^\up-a^\up b^\up c^\dw-a^\up b^\dw c^\up)/
\sqrt{6}$, $abc\chi_A=(a^\up b^\up c^\dw-a^\up b^\dw c^\up)/\sqrt{2}$, and
$(ij)$ means permutation for the quark in place $i$ with the quark in place
$j$. As an example of the bag model evaluation,
we look at the decay $\Xi_b^0\to\Xi_c^0\gamma$. With the
wave functions given by (4.53), we find
$$\la  \Xi_c^0\up|b^\dagger_{1c}b_{1b}b^\dagger_{2d}b_{2u}\sigma_1^z|\Xi_b^0
\up\ra=\,{1\over 6},\eqno(4.54a)$$
$$\la  \Xi_c^0\up|b^\dagger_{1c}b_{1b}b^\dagger_{2d}b_{2u}\sigma_2^z|\Xi_b^0
\up\ra=\,0,\eqno(4.54b)$$
$$\la  \Xi_c^0\up|b^\dagger_{1c}b_{1b}b^\dagger_{2d}b_{2u}(\sigma_{1+}\sigma_{
2-}-\sigma_{1-}\sigma_{2+})|\Xi_b^0\up\ra =\,0,\eqno(4.54c)$$
$$\la  \Xi_c^0\up|b^\dagger_{1c}b_{1u}b^\dagger_{2d}b_{2b}\sigma_1^z|\Xi_b^0
\up\ra=\,{1\over 12},\eqno(4.54d)$$
$$\la  \Xi_c^0\up|b^\dagger_{1c}b_{1u}b^\dagger_{2d}b_{2b}\sigma_2^z|\Xi_b^0
\up\ra=\,{1\over 12},\eqno(4.54e)$$
$$\la  \Xi_c^0\up|b^\dagger_{1c}b_{1u}b^\dagger_{2d}b_{2b}(\sigma_{1+}\sigma_{
2-}-\sigma_{1-}\sigma_{2+})|\Xi_b^0\up\ra =\,{1\over 12}.\eqno(4.54f)$$
For the decay $\Xi_b^0\to\Xi_c^0\gamma$, we find finally
$$\begin{array}{ccl}
a_1 =-b_2 &=& {4\pi\over 3}\int_0^R r^2dr(v_cu_b+u_cv_b)(v_du_u-u_dv_u),   \\
b_1 =-a_2 &=& -{2\pi\over 3}\int_0^R r^2dr[(3u_cu_b-v_cv_b)(u_du_u+v_dv_u)+
(u_cv_b-v_cu_b)(u_dv_u-v_du_u)].  \\
\end{array}\eqno(4.55)$$
Consequently,
$$a_\pm=\,{1\over 2}I_\pm,~~~~b_\pm=\mp {1\over 2}I_\pm,\eqno(4.56)$$
with
$$I_+=\,{4\pi\over 3}\int_0^R r^2dr[(u_cv_b+3v_cu_b)(v_du_u-u_dv_u)+(3u_cu_b-
v_cv_b)(u_uu_d+v_uv_d)],\eqno(4.57a)$$
$$I_-=\,{4\pi\over 3}\int_0^R r^2dr[(3u_cv_b+v_cu_b)(v_du_u-u_dv_u)-(3u_cu_b-
v_cv_b)(u_uu_d+v_uv_d)].\eqno(4.57b)$$
{}From Eqs.(4.50) and (4.56) we obtain
$$a=\,-{G_F\over \sqrt{2}}V_{cb}V_{ud}^*c_-(d_fI_--d_iI_+),\eqno(4.58a)$$
$$b=\,-{G_F\over \sqrt{2}}V_{cb}V_{ud}^*c_-(d_fI_-+d_iI_+).\eqno(4.58b)$$

  Several remarks are in order. (i) We have explicitly confirmed that
the operator $O_+^F$ does not contribute to the baryon transition matrix
elements as the baryon-color wave function is totally antisymmetric. This is
ascribed to the fact that $O_+^F$ is symmetric in color indices, as can be seen
by applying the Fierz transformation to the effective operator for the
amplitude (2.10) and by noting
that the photon interaction is color singlet. From Eq.(4.43) we conclude that
$A=4A'$ and $B=4B'$.
(ii) In the isospin limit we have $I_+=-I_-$, so there is only one independent
bag integral for the decay amplitude of $\Xi_b^0\to \Xi_c^0\gamma$. In the
same limit, we find that
$$\left.{a\over b}\right|_{\Xi_b^0\to\Xi_c^0\gamma}=\,{e_c{m_f\over m_c}-e_d
{m_f\over m_d}
+\left(e_u{m_i\over m_u}-e_b{m_i\over m_b}\right)\over e_c{m_f\over m_c}-e_d
{m_f\over m_d}-\left(e_u{m_i\over m_u}-e_b{m_i\over m_b}\right)}.
\eqno(4.59)$$
Comparing this with (4.44), it appears that the bag model does not predict
correctly the ratio $a/b$. This seeming inconsistency comes from the
static bag approximation we have adopted. In the rest frame of the initial
baryon, one can show that the ratio of the masses $m_i$ and $m_f$ is given
by
$$r\equiv{m_i\over m_f}=\sqrt{1+v_f/c\over 1-v_f/c}.\eqno(4.60)$$
The ratio $a/b$ is in principle a function of the quark masses, the bag
parameters and the velocity $v_f$. But in the static bag approximation, we
always have $r=1$. In order to get the heavy quark symmetry prediction for
$a/b$,
we thus need to utilize a moving bag to describe the recoil effect of the final
baryon state. The net effect should be that (4.58) is modified to
$$a=\,-{G_F\over \sqrt{2}}V_{cb}V_{ud}^*c_-I_-{e\over m_i^2-m_f^2}\left[m_f
\left({e_c\over m_c}-{e_d\over m_d}\right)+m_i\left({e_u\over m_u}-{e_b
\over m_b}\right)\m\right],\eqno(4.61a)$$
$$b=\,-{G_F\over \sqrt{2}}V_{cb}V_{ud}^*c_-I_-{e\over m_i^2-m_f^2}\left[m_f
\left({e_c\over m_c}-{e_d\over m_d}\right)-m_i\left({e_u\over m_u}-{e_b
\over m_b}\right)\m\right],\eqno(4.61b)$$
as implied by Eq.(4.43). Later we will use (4.61) rather than (4.58) to
compute the decay rate and branching ratio for the decay $\Xi_b^0\to\Xi_c^0
\gamma$. (iii) Experimentally, the ratio $a/b$ can be determined
by measuring the asymmetry parameter $\alpha$ when the initial baryon is
polarized with the polarization vector $\vec{s}_i$:
$${d\Gamma(B_i\to B_f\gamma)\over d\Omega}=\,{1\over 4\pi}\Gamma(B_i\to B_f
\gamma)(1+\alpha\vec{s}_i\cdot\hat{p}_f),\eqno(4.62)$$
where
$$\alpha=\,{2{\rm Re}(a^*b)\over |a|^2+|b|^2}.\eqno(4.63)$$

   For completeness, we shall write down the results for the remaining two
decay modes of the bottom baryon. For $\Xi_b^0\to{\Xi'}^0_c\gamma$, we get
$$\begin{array}{ccl}
a_1=-b_2 &=& {4\pi\over 3\sqrt{3}}\int^R_0r^2dr[u_cu_b(3u_uu_d-v_uv_d)+2v_cu_b(
u_uv_d+v_uu_d)-v_cv_b(u_uu_d+v_uv_d)],  \\
b_1=-a_2 &=&-{4\pi\over 6\sqrt{3}}\int^R_0r^2dr[v_cu_b(5u_uv_d-v_uu_d)-u_cv_b(
3v_uu_d+u_uv_d)   \\  &&~+(u_cu_b+v_cv_b)(3u_uu_d-v_uv_d)].
\end{array}\eqno(4.64)$$
As a result,
$$a_\pm=\,{1\over 2}I'_\pm,~~~~b_\pm=\mp {1\over 2}I'_\pm,\eqno(4.65)$$
with
$$I'_+=\,{4\pi\over 3\sqrt{3}}\int^R_0 r^2dr[3(u_cu_d+v_cv_d)(3u_uu_b-v_uv_b)+(
3u_bv_u+v_bu_u)(u_dv_c-v_du_c)],\eqno(4.66a)$$
$$I'_-=\,{4\pi\over 3\sqrt{3}}\int^R_0 r^2dr[(3u_cu_d-v_cv_d)(u_uu_b+v_uv_b)+(
u_bv_u-v_bu_u)(5u_dv_c-v_du_c)].\eqno(4.66b)$$
The resulting amplitudes for the decay $\Xi_b^0\to{\Xi'}_c^0\gamma$ are
$$a=\,-{G_F\over \sqrt{2}}V_{cb}V_{ud}^*c_-(d_fI'_--d_iI'_+),\eqno(4.67a)$$
$$b=\,-{G_F\over \sqrt{2}}V_{cb}V_{ud}^*c_-(d_fI'_-+d_iI'_+).\eqno(4.67b)$$
As for the transition $\Lambda_b^0\to\Sigma_c^0\gamma$, we find that its
amplitude is the same as that of $\Xi_b^0\to{\Xi'}_c^0\gamma$ except for
a different overall normalization factor. More precisely,
$$a(\Lambda_b^0\to\Sigma_c^0\gamma) =\,\sqrt{2}a(\Xi_b^0\to{\Xi'}_c^0\gamma),
\eqno(4.68a)$$
$$b(\Lambda_b^0\to\Sigma_c^0\gamma) =\,\sqrt{2}b(\Xi_b^0\to{\Xi'}_c^0\gamma).
\eqno(4.68b)$$

    Finally, we come to numerical estimates. For the bag parameters we use
[27]
\footnote{It should be stressed that except for the bag quark masses used
in (4.69), all the light quark masses employed in the present paper are of the
constituent type [see Eq.(3.10)], as explained in Sect. II.}
$$\begin{array}{ccl}
&& m_u=m_d=0,~~~m_s=0.279\,{\rm GeV},~~~m_c=1.551\,{\rm GeV},~~~m_b=5.0\,{\rm
GeV},   \\    && x_u=2.043\,,~~~x_s=2.488\,,~~~x_c=2.948\,,~~~x_b=3.079\,,~~~
R=5.0\,{\rm GeV}^{-1},
\end{array}\eqno(4.69)$$
where $R$ is the bag radius.
The eigenvalues $x_q$'s are determined by the transcendental equation
$$\tan x=\,{x\over 1-mR-(x^2+m^2R^2)^{1/2}}.\eqno(4.70)$$
Numerically, the relevant bag integrals are found to be
$$\begin{array}{ccl}
 && I_+=-I_-=\,2.443\times 10^{-3}\,{\rm GeV}^3, \\    &&I'_+=\,3.720\times
10^{-3}\,{\rm GeV}^3,~~~~I'_-=\,1.267\times 10^{-3}\,{\rm GeV}^3.
\end{array}\eqno(4.71)$$
Putting everything together and using $m_{_{\Xi_b}}=5809$ MeV [29],
$m_{_{\Xi'_c}}=2573$ MeV, we finally obtain the decay rates
$$\Gamma(\Xi_b^0\to\Xi_c^0\gamma)=\,3.95\times 10^{-20}\,{\rm GeV},
\eqno(4.72a)$$
$$\Gamma(\Xi_b^0\to{\Xi'}_c^0\gamma)=\,3.54\times 10^{-19}\,{\rm GeV},
\eqno(4.72b)$$
$$\Gamma(\Lambda_b^0\to\Sigma_c^0\gamma)=\,6.65\times 10^{-19}\,{\rm GeV},
\eqno(4.72c)$$
the branching ratios
$${\cal B}(\Xi_b^0\to\Xi_c^0\gamma)=\,6.4\times 10^{-8},\eqno(4.73a)$$
$${\cal B}(\Xi_b^0\to{\Xi'}_c^0\gamma)=\,5.7\times 10^{-7},\eqno(4.73b)$$
$${\cal B}(\Lambda_b^0\to\Sigma_c^0\gamma)=\,1.1\times 10^{-6},
\eqno(4.73c)$$
for $\tau(\Xi_b^0)\sim\tau(\Lambda_b^0)= 1.07\times 10^{-12}s$ [15], and
the decay asymmetry
$$\alpha(\Xi_b^0\to\Xi_c^0\gamma)=\,-0.47\,,\eqno(4.74a)$$
$$\alpha(\Xi_b^0\to{\Xi'}_c^0\gamma)=\,-0.98\,,\eqno(4.74b)$$
$$\alpha(\Lambda_b^0\to\Sigma_c^0\gamma)=\,-0.98\,.\eqno(4.74c)$$
Note that the prediction of $\alpha$ for the decay mode $\Xi_b^0\to\Xi_c^0
\gamma$ is based on heavy quark symmetry [cf. Eq.(4.44)].
Since the baryonic matrix elements are evaluated under the
static bag approximation which amounts to a maximal overlap of wave functions,
the decay rates and branching ratios given by (4.72) and (4.73) for
$\Xi_b^0\to{\Xi'}_c^0\gamma$ and $\Lambda_b^0\to\Sigma_c^0\gamma$ ought to be
regarded as the most optimistic estimates, and
the respective decay asymmetry parameters as order of magnitude estimate.
Nevertheless, the sign of $\alpha$ given in (4.74) is more trustworthy.
Finally, a comparsion of (4.73) with (4.18) leads to the conclusion that the
weak radiative decays of bottom baryons are indeed dominated by the
electromagnetic penguin mechanism.

\vskip 0.5 cm
\noindent{\bf V.~~~Conclusions}
\vskip 0.4cm
    Nonpenguin weak radiative decays of heavy hadrons are characterized by
emission of
a hard photon and the presence of a highly virtual intermediate quark between
the electromagnetic and weak vertices. We have argued that these features
should make possible to analyze these processes by perturbative QCD.

    In this work we have found in tree approximation that these processes
are describable by an effective local and gauge invariant Lagrangian. This
Lagrangian leads to a unique heavy quark symmetry prediction for the asymmetry
parameter in the decay $\Xi^0_b\to\Xi_c^0\gamma$. Other interesting results
are obtained by making use of the effective Lagrangian in conjunction with
the factorization approximation for heavy meson decays and the MIT bag
model for heavy baryon decays. In particular, the branching ratio for
$\bar{B}^0\to D^{0*}\gamma$ is found to be $0.9\times 10^{-6}$. This is very
important for the experimental interpretation of the inclusive measurement of
$\bar{B}^0\to\gamma+$ {\it anything} and its relation to the penguin dominated
decay $\bar{B}^0\to \bar{K}^{0*}\gamma$. In Section 4 we have presented two
different methods for estimating the rates of $\Lambda_b\to\Lambda\gamma$.
Both approaches provide a consistent order of magnitude estimate, namely
${\cal B}(\Lambda_b\to\Lambda\gamma)\sim 1\times 10^{-5}$.
We conclude that weak radiative
decays of bottom hadrons are dominated by the short-distance $b\to s\gamma$
mechanism.

   The factorization method has been known to be reliable for nonleptonic
decays of heavy mesons. So the prediction for $\bar{B}^0\to D^{*0}\gamma$
based on this method should also be reliable.
For the heavy baryon sector, we have to resort to the static bag
approximation even though the initial and final heavy baryons move with
substantially different velocities. This is a serious drawback in our work.
The bag results obtained in Section 4 can only be regarded as order of
magnitude's rough estimates. It remains an important theoretical question how
to incorporate the relative motion between two bags in the MIT bag model.

   Cabibbo-allowed weak radiative decays of charmed hadrons give rise
to strange hadron in the final states. The constitutent $s$ quark, whose mass
is of order 500 MeV, is not very heavy. We can make rough estimates of the
branching ratios for the aforementioned decays by treating the $s$ quark
as heavy and applying the formalism developed in this paper. Even though the
$1/m_s$ corrections are expected to be $(50-100)\%$, they should not alter
the order of magnitudes. In this way, we find
$${\cal B}(D^0\to \bar{K}^{*0}\gamma)=\,1.1\times 10^{-4},\eqno(5.1)$$
$${\cal B}(\Lambda^+_c\to\Sigma^+\gamma)=\,4.9\times 10^{-5},~~~
{\cal B}(\Xi_c^0\to\Xi^0\gamma)=\,3.1\times 10^{-5},\eqno(5.2)$$
$$\alpha(\Lambda^+_c\to\Sigma^+\gamma)=\,-0.86\,,~~~~
\alpha(\Xi_c^0\to\Xi^0\gamma)=\,-0.86\,,\eqno(5.3)$$
where we have used the lifetimes for $D,~\Lambda_c,~\Xi_c$ from PDG [15].
The branching ratio for $D^0\to \bar{K}^{*0}+\gamma$ is quite sizeable. These
decays should be searched for experimentally.

   Finally, we observe that in the weak radiative decays of bottom hadrons
the highly virtual quark's squared invariant mass
is of order $m_b^2$ or smaller. It is therefore appropriate to employ the
renormalization group improved weak interaction Hamiltonian (2.4) with a
renormalization scale $\mu=m_b$ as a starting point. If one wishes to use
a renormalization scale smaller than $m_b$, one must reanalyze the one loop
corrections in a heavy quark effective theory including diagrams with
the photon inside the loop. However, we have not considered such an analysis.

\vskip 2.0 cm
\centerline{\bf Acknowledgments}
\bigskip
T.M.Y.'s work is supported in part by the
National Science Foundation.  This research is supported in part by the
National Science Council of ROC under Contract Nos.  NSC83-0208-M001-012,
NSC83-0208-M001-014, NSC83-0208-M001-015 and NSC83-0208-M008-009.

\pagebreak

\centerline{\bf REFERENCES}

\bigskip
\noindent{$^\dagger$~Address after August 1, 1994: Institute of Physics,
National Chiao Tung University, Hsinchu 30035, Taiwan, ROC}
\begin{enumerate}

\item CLEO Collaboration, R. Ammar {\it et al.,} {\sl Phys. Rev. Lett.}
{\bf 71}, 674 (1993).

\item H.Y. Cheng, C.Y. Cheung, G.L. Lin, Y.C. Lin, T.M. Yan, and H.L. Yu,
{\sl Phys. Rev.} {\bf D47}, 1030 (1993).

\item A.N. Kamal and R.C. Verma, \pr {\bf D26}, 190 (1982).

\item A.N. Kamal and Riazuddin, \pr {\bf D28}, 2317 (1983).

\item L.C. Hua, \pr {\bf D26}, 199 (1982).

\item H.Y. Cheng, ITP-SB-93-41, talk presented at the {\it 5th International
Symposium on Heavy Flavor Physics}, Montreal, July 6-10, 1993.

\item M. Misiak, \np {\bf B393}, 23 (1993); \pl {\bf B269}, 161 (1991); G.
Cella,
G. Curci, G. Ricciardi, and A. Vicer\'e, \pl {\bf B248}, 181 (1990); R.
Grigjanis, P.J. O'Donnell, M. Sutherland, and H. Navelet, \pl {\bf B237}, 355
(1990); B. Grinstein, R. Springer, and M.B. Wise, \np {\bf B339}, 269 (1990).

\item P. Cho and H. Georgi, \pl {\bf B296}, 408 (1992); {\bf 300}, 410(E)
(1993); J.F. Amundson, C.G. Boyd, E. Jenkins, M. Luke, A.V. Manohar, J.L.
Rosner, M.J. Savage, and M.B. Wise, {\sl ibid} {\bf B296}, 415 (1992).

\item For a review, see H.Y. Cheng, {\sl Int. J. Mod. Phys.} {\bf A4}, 495
(1989).

\item L.L. Chau, H.Y. Cheng, W.K. Sze, and H. Yao, \pr {\bf D43}, 2176 (1991).

\item A. Deandrea, N. Di Bartolomeo, R. Gatto, and G. Nardulli,
\pl {\bf B318}, 549 (1993); K. Honscheid, K.R. Schubert, and R. Waldi,
OHSTPY-HEP-E-93-017 (1993); S. Resag and M. Beyer, BONN-TK-93-18 (1993).

\item S. Stone, HEPSY 93-4, talk presented at the {\it 5th International
Symposium on Heavy Flavor Physics}, Montreal, July 6-10, 1993.

\item CLEO Collaboration, M.S. Alam {\it et al.,} CLNS 84-1270 (1994).

\item R.R. Mendel and P. Stiarski, \pr {\bf D36}, 953 (1987).

\item Particle Data Group, \pr {\bf D50}, 1173 (1994).

\item M. Neubert, CERN-TH-7395/94 (1994).

\item P. Colangelo, G. Nardulli, N. Paver, and Riazuddin, \zp {\bf C45}, 575
(1990).

\item T.M. Yan, H.Y. Cheng, C.Y. Cheung, G.L. Lin, Y.C. Lin, and H.L. Yu,
{\sl Phys. Rev.} {\bf D46}, 1148 (1992); see also T.M. Yan, {\sl Chin. J.
Phys.} (Taipei) {\bf 30}, 509 (1992).

\item H. Georgi, B. Grinstein, and M.B. Wise, \pl {\bf B252}, 456 (1990).

\item A. Falk {\it et al.}, \np {\bf B343}, 1 (1990).

\item E. Jenkins, A. Manohar, and M.B. Wise, \np {\bf B396}, 38 (1993).

\item M. Sadzikowski and K. Zalewski, {\sl Z. Phys.} {\bf C59}, 677 (1993).

\item T. Mannel, W. Roberts, and Z. Ryzak, \np {\bf B355}, 38 (1991); \pl
{\bf B255}, 593 (1991); F. Hussain, J.G. K\"orner, M. Kr\"amer, and G.
Thompson, \zp {\bf C51}, 321 (1991).

\item R. P\'erez-Marcial, R. Huerta, A. Garcia, and M. Avila-Aoki, \pr
{\bf D40}, 2955 (1989).

\item CLEO Collaboration, J. Dominick {\it et al.,} CLEO CONF 94-19 (1994).

\item M. Avila-Aoki, A. Garcia, R. Heurta, and R. P\'erez-Marcial, \pr
{\bf D40}, 2944 (1989).

\item A. Chodos, R.L. Jaffe, K. Johnson, and C.B. Thorn, \pr {\bf D10}, 2599
(1974); T. DeGrand, R.L. Jaffe, K. Johnson, and J. Kisis, {\sl ibid} {\bf 12},
2060 (1975).

\item H.Y. Cheng and B. Tseng, \pr {\bf D46}, 1042 (1992) and references cited
therein.

\item U. Aglietti, \pl {\bf B281}, 341 (1992) and hep-ph/9311314.

\end{enumerate}

\pagebreak

\centerline{\bf Figure Captions}

\bigskip

\begin{description}

\item{Fig. 1.}~~$W$-exchange diagrams contributing to the quark-quark
bremsstrahlung process $b+\bar{d}\to c+\bar{u}+\gamma$ induced by the
four-quark operator $O_A$ defined in Eq.(2.6).
\item{Fig. 2.}~~Same as Fig. 1 except for the operator $O_B$ defined in
Eq.(2.6).
\item{Fig. 3.}~~Possible pole diagrams contributing to the radiative decay
$\bar{B}^0\to D^{*0}\gamma$.

\end{description}

\end{document}